\documentclass[aps, prx, twocolumn]{revtex4-2}

\usepackage{graphicx}
\usepackage{dcolumn}
\usepackage{amsmath}
\usepackage{array}
\usepackage[caption=false]{subfig}
\usepackage[bookmarks=false, colorlinks=true, citecolor={blue!60!black}, urlcolor={blue!60!black}, linkcolor = {blue!50!black}]{hyperref}
\usepackage{exscale}
\usepackage[usenames, dvipsnames]{xcolor}
\usepackage[]{cleveref}
\usepackage{microtype}
\usepackage{siunitx}

\Crefformat{appendix}{Appendix~#2#1#3}
\Crefname{section}{Sec.}{Secs.}
\Crefname{subsection}{Sec.}{Secs.}
\Crefname{appendix}{Appendix}{Appendices}
\Crefname{equation}{Eq.}{Eqs.}
\Crefname{figure}{Fig.}{Figs.}
\newcommand{\Longfigref}[1]{Figure~\labelcref{#1}}
\Crefname{tabular}{Tab.}{Tabs.}

\DeclareSIUnit{\ms}{\milli\second}
\DeclareSIUnit{\us}{\micro\second}
\DeclareSIUnit{\ps}{\pico\second}
\DeclareSIUnit{\kHz}{\kilo\hertz}
\DeclareSIUnit{\um}{\micro\meter}
\DeclareSIUnit{\nm}{\nano\meter}
\DeclareSIUnit{\cm}{\centi\meter}
\DeclareSIUnit{\mV}{\milli\volt}
\DeclareSIUnit{\uV}{\micro\volt}
\DeclareSIUnit{\ueV}{\micro\electronvolt}
\DeclareSIUnit{\meV}{\milli\electronvolt}
\DeclareSIUnit{\mT}{\milli\tesla}
\DeclareSIUnit{\mK}{\milli\kelvin}
\newcommand{\Q}{\mathrm{\scriptscriptstyle Q}}
\newcommand{\Cq}{C_\Q}
\newcommand{\Cqh}{C_{\Q\mathrm{h}}}
\newcommand{\Cql}{C_{\Q\mathrm{l}}}

\newcommand{\X}{{\scriptscriptstyle X}}
\newcommand{\Z}{{\scriptscriptstyle Z}}

\newcommand{\tauX}{\tau_\X}
\newcommand{\tauZ}{\tau_\Z}

\newcommand{\DeltaT}{\Delta_\mathrm{\scriptscriptstyle T}}
\newcommand{\xiT}{\xi_\mathrm{\scriptscriptstyle T}}

\newcommand{\EM}{E_\mathrm{\scriptscriptstyle M}}

\newcommand{\EP}{E_{\scriptscriptstyle P}}
\newcommand{\tP}{t_{\scriptscriptstyle P}}

\newcommand{\SA}{S_\mathrm{\scriptscriptstyle A}}
\newcommand{\SB}{S_\mathrm{\scriptscriptstyle B}}

\newcommand{\Vp}{V_\mathrm{p}}
\newcommand{\Vwp}{V_\mathrm{wp}}

\renewcommand{\L}{\mathrm{\scriptscriptstyle L}}
\newcommand{\R}{\mathrm{\scriptscriptstyle R}}

\newcommand{\kB}{k_\mathrm{\scriptscriptstyle B}}
\newcommand{\vF}{v_\mathrm{\scriptscriptstyle F}}
\newcommand{\Tc}{T_\mathrm{\scriptscriptstyle C}}

\begin{document}

\title{20 Second Parity Lifetime in an InAs--Pb Tetron Device}

\author{Microsoft Quantum$^\dagger$}
\noaffiliation

\date{\today}

\begin{abstract}
A central promise of topological quantum computing is that increasing the excitation gap improves device performance significantly.
Here, we experimentally validate this principle in an InAs--Pb tetron device via interferometric single-shot parity measurements.
By replacing aluminum with the higher-gap superconductor lead in our superconductor-semiconductor hybrid devices, we have improved the robustness of our topological phase.
In addition, to enable fast and precise bring-up at scale, we have developed an rf measurement technique that resolves low-energy wire-end states and directly measures their energy splitting with \si{\ueV} precision.
We employ this technique to bring up a device in a multi-tetron array and perform parity measurements of one of the tetron's hybrid nanowires (NWs).
By controllably switching the wire parity, we observe $h/2e$-periodic bimodal shifts in the quantum capacitance of a quantum dot coupled to the hybrid nanowire in an interference loop.
Further time-resolved measurements reveal a characteristic parity switching time of $\sim \SI{20}{\second}$ with some instances reaching minute-scale.
Such extremely long parity lifetimes are orders of magnitude longer than typical qubit operation times, which are on the order of \si{\us}.
Finally, we discuss potential implications for the fidelity of Pauli measurements.
\end{abstract}

\maketitle

\section{Introduction}
\label{sec:introduction}

Recently, we presented a roadmap~\cite{Aasen25} to fault-tolerant quantum computation using topological qubits~\cite{Kitaev97, Freedman98, Nayak08} built around Majorana zero modes (MZMs) in superconductor-semiconductor hybrid devices~\cite{Kitaev01, Fu08, Sau10a, Lutchyn10, Oreg10}.
Our roadmap draws on concepts explored in Refs.~\onlinecite{Fu09, Akhmerov09, Fu09a, Fu10, Hassler10, Sau11a, Fidkowski11b, Alicea12a, DasSarma15, Heck11, Heck12, Hyart13, Houzet13, Pientka13a, Cheng15, Sau15, Yavilberg15, Aasen16, Hell18, Chiu18, Drukier18, Vijay15, Vijay16a, Vijay16b, Knapp18a, Knapp18b, Liu19, Munk20, Steiner20, Khindanov21a}.
It entails building successively larger arrays of qubits which are progressively deeper inside the topological phase, characterized by larger energy gap and shorter coherence length.
In superconductor-semiconductor hybrid platforms, these key parameters are governed primarily by the parent superconducting gap, the semiconductor spin-orbit coupling, and material disorder~\cite{Motrunich01, Gruzberg05, Brouwer11a, Brouwer11b, Stanescu11, Lobos12, Sau12, DeGottardi12, DeGottardi13, Pekerten17, DasSarma21, Ahn21, DasSarma23}.
In this work, we demonstrate that engineering these ingredients---by replacing Al with the higher-gap superconductor Pb and enhancing spin-orbit coupling via band-structure design in a semiconductor heterostructure grown on a GaSb substrate---leads to substantial improvements in device performance.
By demonstrating these performance improvements, we have gone beyond previous work which has incorporated larger gap superconductors into nanowire devices \cite{Kanne21, Pendharkar19, Song25, Zhang26}.

The building block of our architecture is a tetron \cite{Plugge17, Karzig17}, 
which is composed of two parallel proximitized semiconductor NWs connected by a conventional superconducting backbone, forming an H-shaped superconducting island.
Each nanowire hosts a pair of Majorana zero modes (MZMs) at its ends when tuned into the topological phase, yielding four MZMs per tetron, see also \cite{Aghaee25b}.
The computational subspace has fixed total electron parity, and projective Pauli measurements of the computational qubit are implemented by measuring the parity of different pairs of MZMs in the tetron.
The dominant error mechanisms arise from (i)~quasiparticle poisoning, where an MZM exchanges a fermion with the continuum above the topological gap $\DeltaT$, and (ii)~Majorana hybridization, in which tunneling processes between MZMs generate a finite energy splitting $\EM$.

To reliably access the topological regime, we developed the topological gap protocol (TGP) \cite{Aghaee23}, which identifies the topological phase with high fidelity and reports the topological gap $\DeltaT$.
The TGP was used to tune our devices into the qubit configuration.
In earlier Al-based devices, we observed typical top quintile gaps of $\DeltaT \sim \SI{30}{\ueV}$~\cite{Aghaee23, Aghaee25a, Aghaee25b}.
In contrast, in the InAs--Pb devices studied here we observe a top quintile gap of $\DeltaT \sim \SI{70}{\ueV}$.
In parallel, we introduce a direct probe of Majorana hybridization, in which a quantum dot (QD) measures one wire end while a QD at the opposite end is used to controllably change fermion parity in the nanowire.
This technique reveals extended parameter regimes consistent with  $\EM < \SI{1}{\ueV}$, i.e. smaller than our experimental resolution.

Topological protection gives error rates that are exponentially suppressed as the topological gap increases~\cite{Knapp18a}.
This expectation, however, relies critically on the assumption that the system remains near equilibrium.
In the presence of non-equilibrium processes, this protection can be significantly degraded.
A large topological gap $\DeltaT$ suppresses thermally-activated excitations, but it does not protect against far-from-equilibrium quasiparticles~\cite{Aumentado04, Naaman06, Lutchyn06, Ferguson06, Shaw08, Persson10, Barends11, Riste13, Janvier15, Hays18, Serniak19a, Uilhoorn21, Mannila22, Erlandsson23, Wesdorp23}.
Such quasiparticles can be generated, for example, when residual radiation penetrates shielding and breaks Cooper pairs, and persist if recombination processes are slow.
As a result, the quasiparticle poisoning rate and associated density of non-equilibrium quasiparticles emerges as a key metric, distinct from both $\DeltaT$ and the Majorana splitting $\EM$.
In this work, we show that the quasiparticle poisoning rate is also strongly suppressed in InAs--Pb devices, complementing the increase of $\DeltaT$ and further improving device robustness.

Qubit operations in our architecture are realized via quantum-capacitance readout of QDs tunnel-coupled to MZMs.
For example, a Pauli-$Z$ measurement is implemented by probing a QD coupled to both ends of a single topological wire.
In such measurements of Al-based tetrons, we previously measured parity lifetimes of $\sim \SIrange{1}{12}{\ms}$ \cite{Aghaee25a, Aghaee25b}.
In contrast, Pb-based devices exhibit $Z$-parity lifetimes exceeding $\SI{20}{\second}$, representing an improvement of more than three orders of magnitude.
In a tetron, the qubit is encoded in the parity, so parity lifetimes directly translate into qubit lifetimes and parity flips are key qubit error processes. 
Consequently, the observed dramatic enhancement in parity stability indicates that non-equilibrium quasiparticles no longer limit qubit operations in our devices, providing strong evidence that increasing superconducting gap directly translates into improved robustness of Majorana-based qubits.

\section{Device design and Pb-based material platform}
\label{sec:device_design}

\begin{figure}
\includegraphics[width=8.64cm]{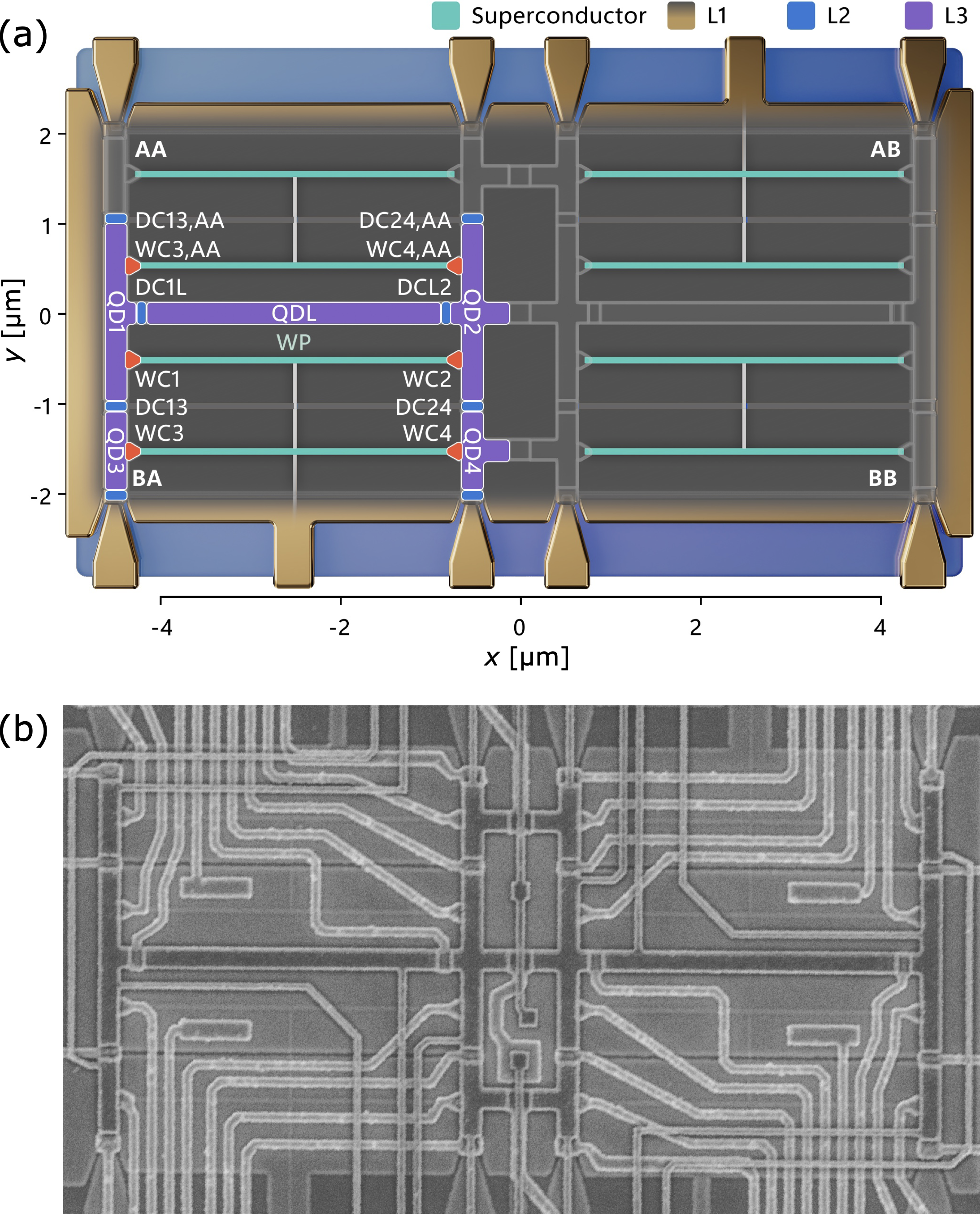}
\caption{
\textbf{Scalable unit cell for multi-qubit tetron array.} 
(a)~Schematic of the tetron-based unit cell used for multi-qubit devices. 
Each tetron consists of two parallel proximitized NWs connected by a superconducting backbone, forming an H-shaped island that hosts four Majorana zero modes (MZMs) at the wire ends when tuned into the topological regime. 
Color-coded gates illustrate the three-layer architecture: electrostatic tuning gates set the nanowire density, junction and cutter gates control tunnel couplings to define and reconfigure interferometric loops, and gate-defined quantum dots (QDs) provide dispersive parity readout. 
Neighboring tetrons are coupled via shared QDs, enabling both single- and two-qubit parity measurements. 
(b)~Scanning electron micrograph (SEM) of a fabricated prototype unit cell for a multi-tetron array, demonstrating a geometry and gate layout compatible with scaling to larger qubit arrays.
}
\label{fig:device}
\end{figure}

The device shown in \Cref{fig:device} is a prototype unit cell for multi-tetron devices.
The qubits are interconnected through shared QDs and tunable junctions.
\Longfigref{fig:device}(a) is a schematic of the device; a scanning electron micrograph (SEM) image is shown in \Cref{fig:device}(b).
This architecture extends earlier single-tetron designs by embedding qubits within a network geometry that enables scalable integration, while supporting both single- and multi-qubit parity measurements via local interactions.

Each tetron consists of two horizontal superconducting NWs of length $\SI{3.5}{\um}$ and width $\SI{35}{\nm}$, connected by a narrower vertical backbone of length $\SI{1}{\um}$ and width $\SI{20}{\nm}$.
This geometry is chosen to simultaneously satisfy several constraints: (a) The wire length exceeds the expected coherence length, suppressing Majorana hybridization and minimizing the splitting energy $\EM$.
(b) The narrower backbone enables full depletion when the horizontal wires are tuned to the lowest electronic subband, ensuring that it remains in a trivial state even when the NWs are tuned into the topological phase.
At the same time, it is wide enough to maintain superconducting connectivity.

\begin{figure}
\includegraphics[width=8.64cm]{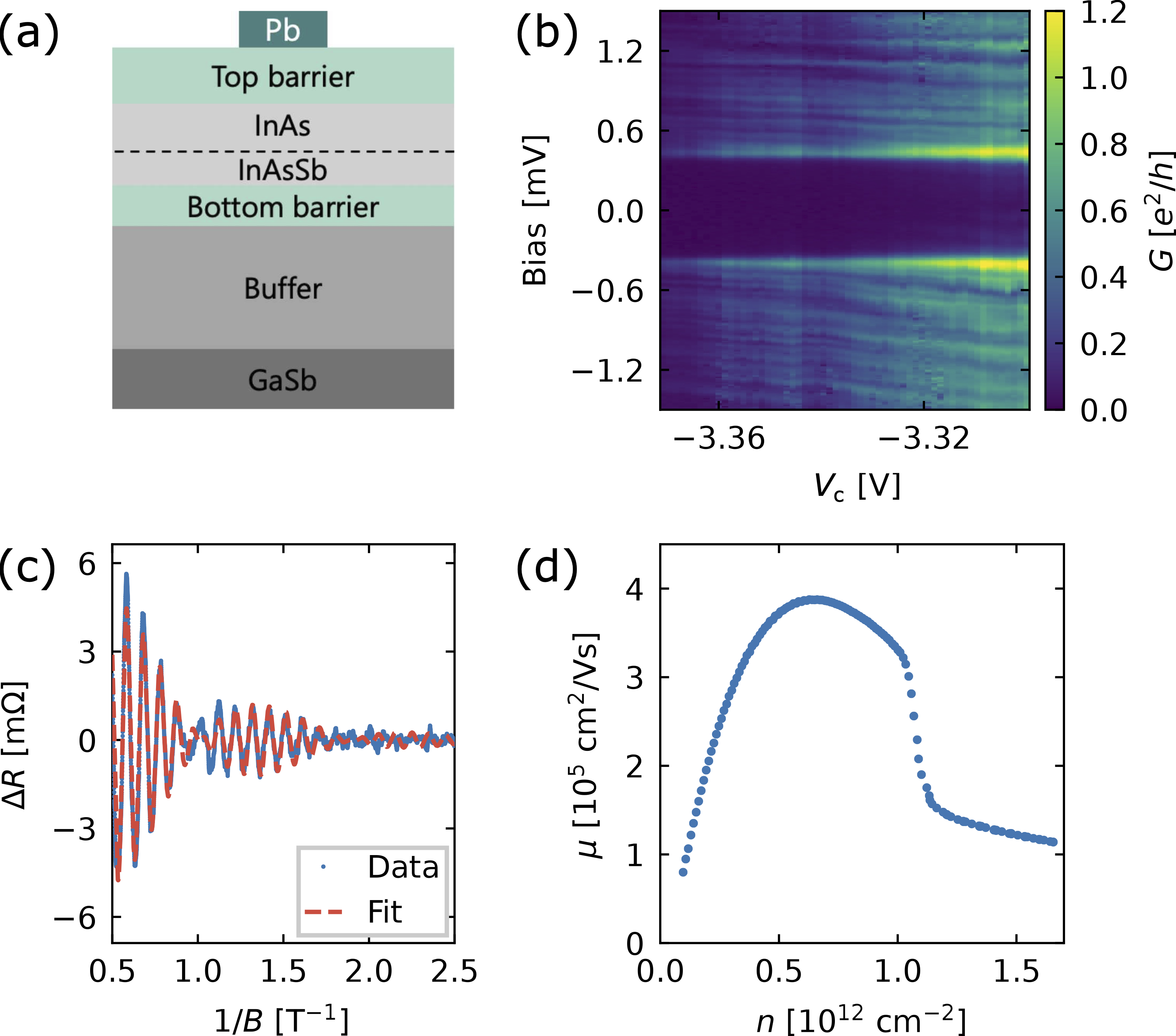}
\caption{
\textbf{Enhanced material stack.}
(a)~Schematic of a hybrid material stack combining Pb with a composite quantum well.
(b)~Conductance as a function of bias voltage and split-gate voltage $V_\mathrm{c}$ in a superconducting quantum point contact to a 2D Pb-covered half-plane in the tunneling regime. A large induced gap is clearly visible. 
(c)~Shubnikov--de Haas oscillations, showing long quantum lifetime and large spin-orbit coupling in a shallow quantum well under Pb, as obtained via the analysis procedure of Ref.~\onlinecite{Zimmerman26}.
We extract a density of $\SI{0.49(2)e12}{\per\cm\squared}$, a spin-orbit coupling of $\SI{12(2)}{\meV\nm}$, and a quantum lifetime of $\SI{0.37(14)}{\ps}$.
(d)~Mobility as a function of density in a buried quantum well that is the analog of the upper part of our semiconductor stack (with the same composite quantum well, top and bottom barrier) used in tetron devices.
}
\label{fig:material_stack1} 
\end{figure}

The device is realized using an epitaxial superconductor-semiconductor heterostructure optimized to enhance the robustness of the topological phase.
A schematic cross-section of our superconductor-semiconductor heterostructure is shown in \Cref{fig:material_stack1}(a).
The H-shaped superconducting structure is formed by depositing $\SI{10}{\nm}$ of Pb on the surface of the semiconductor heterostructure grown on a GaSb substrate.
For this thickness, the parent superconducting gap is $\Delta_\mathrm{\scriptscriptstyle Pb} \approx \SI{1.3}{\meV}$, resulting in a large proximity-induced gap in the semiconductor.
As shown in \Cref{fig:material_stack1}(b), a proximity-induced gap of $\SI{400}{\ueV}$ is measured in a two-dimensional superconductor-quantum point contact geometry.
In the nanowire geometry, the induced gap is increased due to transverse confinement.
We measure an induced gap $\Delta_{\mathrm{ind}} \approx \SI{570}{\ueV}$ in the lowest subband regime at zero magnetic field, consistent with simulations.
For comparison, prior work has shown that Pb-based Coulomb islands exhibit a robust 2$e$-periodic charging pattern at zero magnetic field~\cite{Kanne21}, consistent with the absence of subgap states (i.e., consistent with a ``hard'' induced gap).

The transition to a GaSb substrate provides two principal advantages: (i)~lattice-matched epitaxial growth through the entire semiconductor stack which drastically reduces the extended defect count and improves surface morphology and (ii)~the larger range of accessible band offsets enables larger spin-orbit coupling than is achievable on InP substrates.
The composite semiconductor quantum well consists of a $\SI{6}{\nm}$ InAs layer with an additional $\SI{2}{\nm}$ InAs$_{0.8}$Sb$_{0.2}$ layer, which enhances the Rashba spin-orbit coupling $\alpha \sim \SIrange{12}{16}{\meV\nm}$ in Pb-proximitized NWs [see \Cref{fig:material_stack1}(c)] 
\footnote{
This value is estimated from 2D spin-orbit coupling values obtained from Shubnikov--de Haas oscillations \cite{Zimmerman26} in van der Pauw devices in which there is a Pb layer above the quantum well but the density is fixed, together with weak anti-localization measurements in Hall bars in which the density can be varied but there is no Pb layer.
We use this spin-orbit coupling in our simulations, but the topological gap, $\EM$, and parity lifetime are measured directly and do not depend on this value.
}.
The top barrier thickness and composition are optimized to maintain a large induced gap while avoiding excessive renormalization of the semiconductor $g$-factor and spin-orbit coupling.

\begin{figure}
\includegraphics[width=8.64cm]{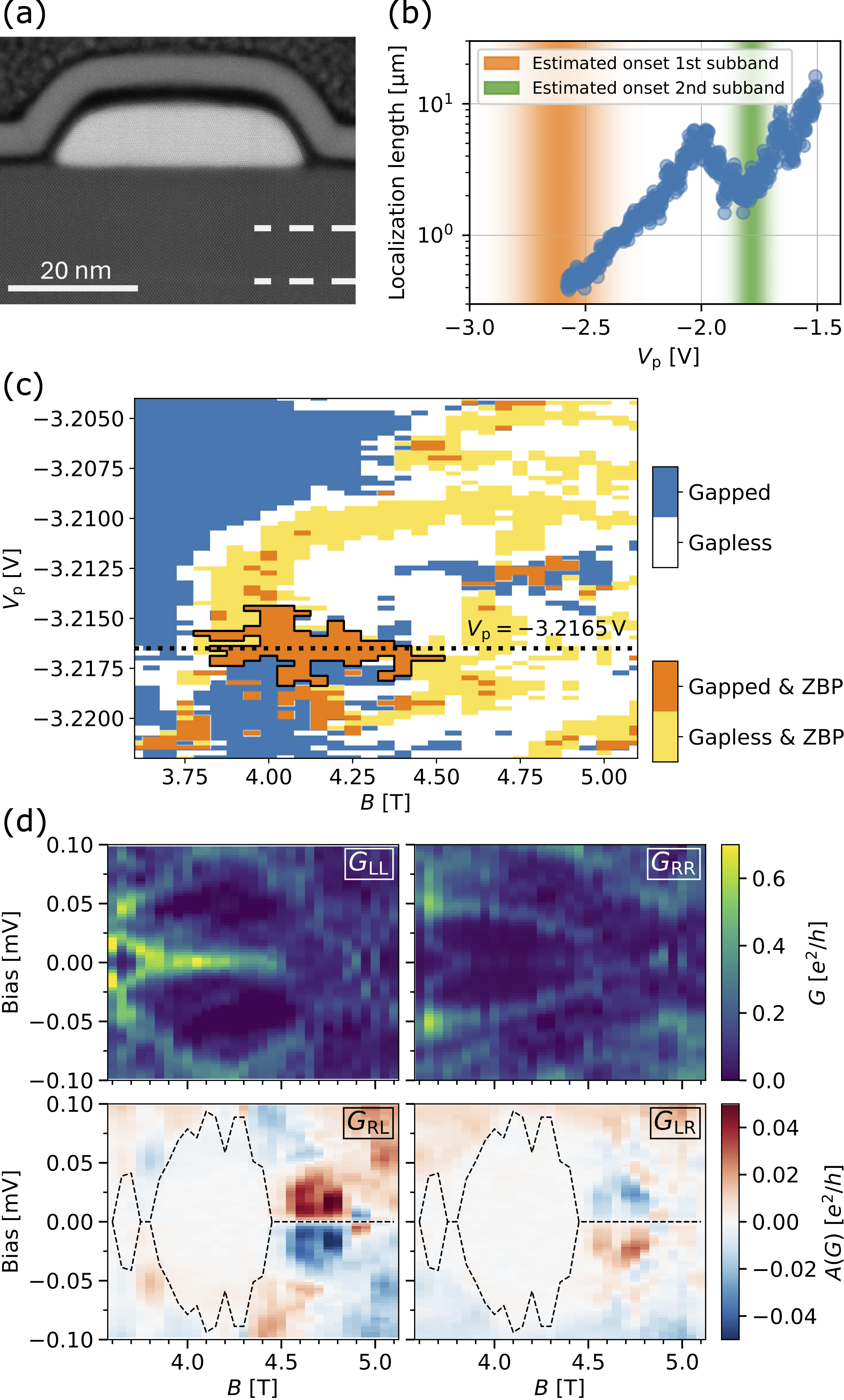}
\caption{
(a)~Cross-sectional TEM of a narrow Pb strip on the surface of semiconductor barrier and quantum well, dashed lines indicate interfaces at the top and bottom of the quantum well. 
(b)~Measured localization length as a function of gate voltage extracted from 14 wires with different lengths of $\SI{1.25}{\um}$ and $\SI{3}{\um}$.
The estimated 1D subband onsets are extracted from quantized local conductance steps.
(c)~Phase diagram obtained from a TGP analysis of transport data from a $\SI{3}{\um}$-long nanowire test structure.
(d)~$B$-field line cut along $\Vp = \SI{3.2165}{\volt}$ showing local and anti-symmetrized non-local conductance along the dashed line in panel (c).
Zero-bias peaks at the opposite ends of the NW persist over the $B$-field range of over $\SI{0.5}{\tesla}$, consistent with gap closing and re-opening.
The dashed (black) line corresponds to extracted gap.
}
\label{fig:material_stack2}
\end{figure}

Disorder is minimized through careful materials engineering of both the epitaxial hybrid superconductor-semiconductor layers and dielectrics.
Measurements of shallow quantum well Hall bar devices indicate a surface charge density $n_\mathrm{\scriptscriptstyle 2D} \approx \SI{2e12}{\per\cm\squared}$.
Buried quantum well Hall bars with the same active region exhibit mobilities $> \SI{350000}{\cm\squared / \volt\second}$ [see \Cref{fig:material_stack1}(d)].
We also observe Shubnikov--de Haas oscillations of electrons residing in the quantum well coupled to the epitaxial metal in van der Pauw devices~\cite{Zimmerman26}.
These results attest to the high material quality of our hybrid stack, with minimal impurity or defect scattering.

The combination of a large induced gap, strong spin-orbit coupling, and low disorder establishes a robust superconducting platform for topological qubits.
Direct evidence for the improved materials quality and its impact on the topological phase is shown in \Cref{fig:material_stack2}.
\Longfigref{fig:material_stack2}(a) presents a cross-sectional transmission electron micrograph (TEM) of a $\SI{35}{\nm}$-wide Pb wire on the surface of the semiconductor barrier and quantum well grown on a GaSb substrate.
The interfaces at the top and bottom of the InAs-based quantum well (indicated by dashed lines) are sharp, and the superconductor-semiconductor interface is likewise abrupt, confirming excellent epitaxial and interfacial quality.

To quantify disorder in Pb-proximitized NWs, we measured the scaling of nonlocal conductance with segment length using the methodology developed in Ref.~\onlinecite{Aghaee23}.
This analysis yields localization lengths exceeding $\SI{1}{\um}$ in the lowest occupied subband, as shown in \Cref{fig:material_stack2}(b).
Such localization lengths are comparable to the full device dimensions and are consistent with the low disorder inferred independently from mobility measurements and Shubnikov--de Haas oscillations.
Importantly, the localization length comfortably exceeds the minimum topological coherence length expected for a clean system ($\xi_\mathrm{cl} \sim \SI{100}{\nm}$), indicating that the topological phase is robust against disorder in these NWs.

Consistent with this conclusion, we observe a substantial improvement over the Al--InAs platform~\cite{Aghaee23} in both the magnitude of the topological gap and the extent of the topological phase in the $(\Vp,B)$ parameter space.
A representative phase diagram obtained from a NW test device is shown in \Cref{fig:material_stack2}(c), which was obtained by applying the Topological Gap Protocol (TGP)~\cite{Aghaee23}.
The outlined orange region denotes the region in wireplunger-field $(\Vp, B)$ parameter space identified as topological.
The region passing TGP exceeds $\SI{1.1}{\mV\tesla}$, more than twice that observed Al--InAs platform with comparable lever arms and $g$ factors~\cite{Aghaee23}.
Within this region, the top-quintile topological gap reaches $\DeltaT \approx \SI{70}{\ueV}$.
Local conductance measurements reveal robust zero-bias peaks at opposite ends of the NW persisting over magnetic-field ranges exceeding $\SI{0.5}{\tesla}$, as seen in \Cref{fig:material_stack2}(d).
These stable zero-bias features, coincident with gap closing and reopening in the nonlocal conductance [\Cref{fig:material_stack2}(d)], constitute hallmark signatures of the topological superconducting phase.
These observations are consistent with our simulations and demonstrate that the improved materials properties directly translate into enhanced topological properties of measured NW test devices, and result in improved qubit performance, as discussed below.

Finally, we outline the key elements of the multi-qubit device design.
The scalable tetron unit cell that underlies our multi-qubit architecture is shown in \Cref{fig:device}.
This unit cell is designed to tile naturally into larger arrays, enabling straightforward extensions to devices with many qubits (e.g., a 12-qubit array) without qualitative changes to the control or readout strategy.
The device employs three distinct gate layers, illustrated in \Cref{fig:device}(a), each serving a well-defined and largely independent function.
The first layer comprises electrostatic tuning gates that control the carrier density in the NWs and provide access to the topological regime.
The second layer consists of junction and cutter gates that tune tunnel couplings, allowing NWs and QDs to be selectively isolated or connected to form configurable interferometric loops for parity measurements.
The third layer enables high-fidelity readout using gate-defined QDs that are dispersively coupled to resonators; the resulting quantum capacitance response provides a sensitive, non-invasive probe of fermion parity \cite{Colless13}.
This layered gate architecture cleanly separates tuning, coupling, and measurement functionalities, while remaining fully compatible with scalable multi-qubit operation.

Two of the four tetron qubits shown in the figure are grounded (AB and BA) and two are floating (AA and BB).
We have incorporated the latter because at least one superconducting island must have charging energy in a two-qubit measurement.
Each H-shaped superconducting island is coupled to five QDs, three of which are shared between vertical neighbors, which can thereby couple.
Concretely, consider the BA qubit.
A Pauli-$Z$ measurement is performed by coupling the top wire in the qubit to quantum dot QDL via QD1 and QD2 and by decoupling these QDs from the AA qubit by using the cutter gates $V_{\mathrm{wc3,\scriptscriptstyle AA}}$, $V_{\mathrm{wc4,\scriptscriptstyle AA}}$, $V_{\mathrm{dc13,\scriptscriptstyle AA}}$, and $V_{\mathrm{dc24,\scriptscriptstyle AA}}$ to close the junctions through which they would couple to it.
In this configuration, QDL's quantum capacitance is sensitive to the fermion parity of the top wire in the BA qubit.
Pauli-$X$ measurements and two-qubit measurements can be performed similarly.

All of the QDs have been designed to have plunger lever arms in the range $\SIrange{0.4}{0.45}{\meV/\mV}$, charging energies $> \SI{60}{\ueV}$, and level spacings $> \SI{30}{\ueV}$.
The long QDs (QDL) are approximately $\SI{3.2}{\um}$ long; the combined linear gate extent of the QDs that couple vertically-neighboring tetrons is $\SI{2.1}{\um}$ (exterior, QD1) or $\SI{2.4}{\um}$ (interior, QD2); while that of the smallest QDs is $\SI{0.9}{\um}$ (exterior, QD3) or $\SI{1.2}{\um}$ (interior, QD4).
These QD parameters have been chosen to facilitate control and readout of the device.

The tetron array is operated in a cryogenic setup with readout and control systems similar to those described in Refs.~\onlinecite{Aghaee25a, Aghaee25b}.

\section{rf-based wire spectroscopy and tuning}
\label{sec:spectroscopy}

While local and non-local DC transport measurements have underpinned the tuning of topological NWs in our earlier work~\cite{Aghaee23, Aghaee25a, Aghaee25b}, they do not offer a scalable characterization strategy for large Majorana-based qubit arrays because they require end-to-end transport paths and a grounded third terminal, thereby limiting parallelization, slowing device bring-up.
More broadly, transport-based protocols are not naturally matched to scalable architectures based on floating hybrid superconductor-semiconductor islands.
A fully rf-based tune-up overcomes these bottlenecks because it is inherently local, compatible with dispersive gate-sensing hardware, and amenable to automation and parallel execution across many qubits.
Moreover, we expect to typically have $\DeltaT / \kB T > 10$ with our new material platform.
At this point, the limiting factor for qubit performance is the residual Majorana energy splitting $\EM$ due to finite wire length rather than thermal excitations above the gap.
As we will show below, rf-based measurements enable higher precision measurement of $\EM$ than DC transport techniques.
In addition, in regimes where quasiparticle-poisoning rates are slow and intrinsic parity-switching becomes inefficient to resolve $\EM$~\cite{Aghaee25a, Aghaee25b}, the parity injection method introduced below allows for a more controllable extraction of $\EM$.

\begin{figure}
\includegraphics[width=8.64cm]{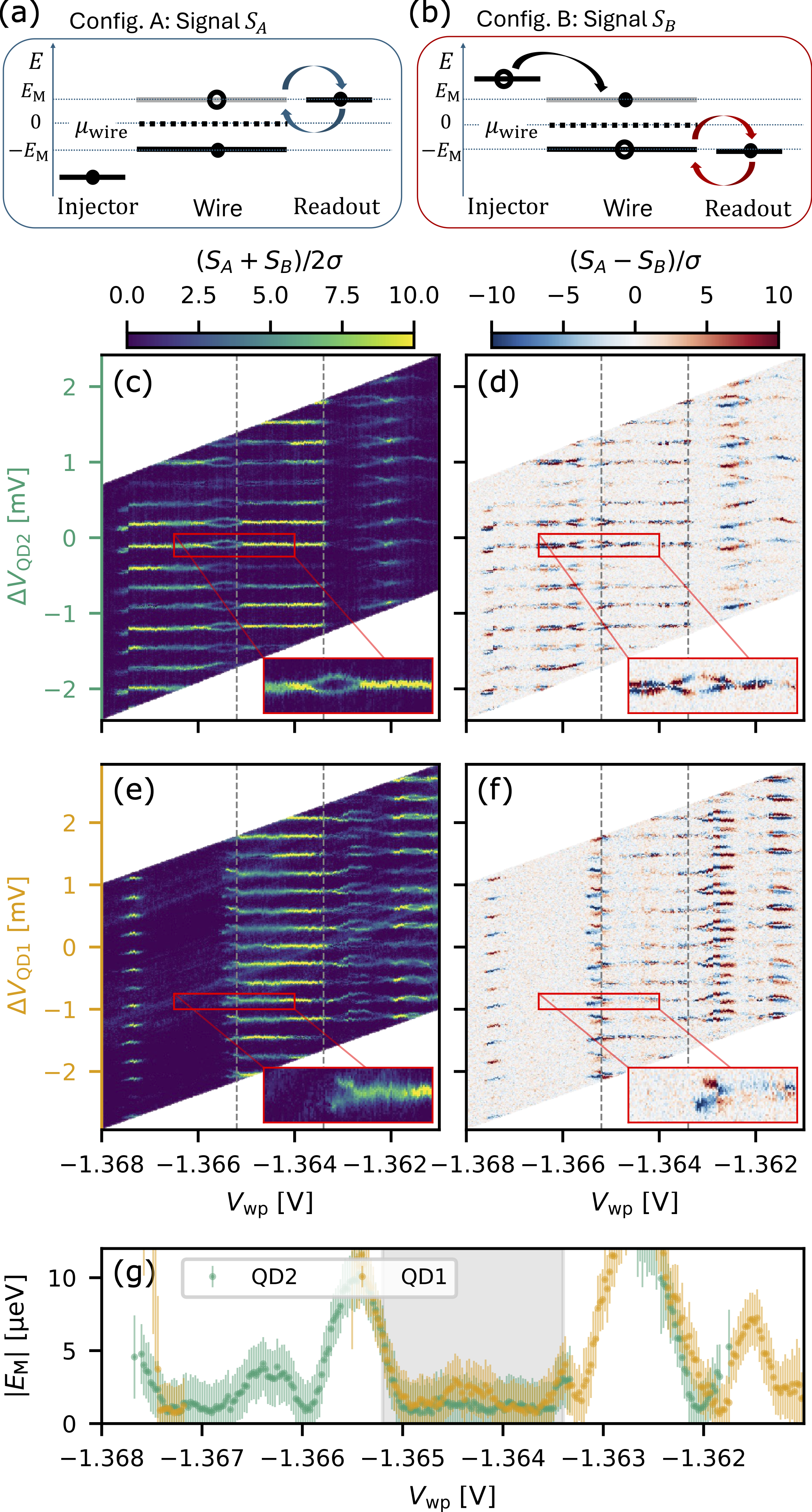}
\caption{
\textbf{Wire spectroscopy.}
(a,b)~Schematic of the quasiparticle injection procedure. 
The injector QD leads to a parity switch of the wire state, which, for finite $\EM$, shifts the wire-readout QD resonance condition.
(c,e)~Mean of the normalized rf signal measured on QD1 (QD2) averaged over QD2 (QD1) configurations where $S$ is the measured rf signal along the quadrature with maximal signal and $\sigma$ the noise along the orthogonal quadrature.
(d,f)~Difference of the normalized signal measured in configurations (a) and (b).
For better visibility, the insets in panels (c--f) each show a zoom-in on parts of a single QD-wire transition.
(g)~Extracted wire state energy $\EM$ based on analyzing the parity-dependent shift of the rf response for all charge transitions.
The gray shaded area indicates the region of interest identified in \Cref{fig:roi}.
To convert from a shift in voltages to detuning energy, we consider a lever arm $\alpha_\mathrm{\scriptscriptstyle QD} = \SI{0.45}{\meV/\mV}$ based on electrostatic simulations.
The axis $\Delta V_\mathrm{\scriptscriptstyle QD1(2)}$ indicate a change in the voltage of the corresponding QD plunger gate where a linear shift is applied to the bare voltage to account for the finite lever arm of the wire plunger on the corresponding QD.
}
\label{fig:em_spectroscopy}
\end{figure}

We focus on the top NW of the BA tetron coupled at both ends to QDs that are in turn connected to resonators for dispersive gate sensing, allowing the quantum capacitance signal $\Cq$ to be measured locally.
As schematically illustrated in \Cref{fig:em_spectroscopy}(a,b), we introduce a spectroscopy measurement where one QD acts as a local probe (readout) and the other acts as a (non-local) parity injector.

To understand the physics behind the measurement, we consider a simple model consisting of a single low-energy wire state with energy $\EM$ coupled via tunneling matrix elements $t_\R$, $\tilde{t}_\R$ to a single level of the readout QD which is detuned by $\delta_\R$.
This model was proposed and explored in Refs.~\onlinecite{Clarke17, Prada17} and used to analyze DC transport data in Ref.~\onlinecite{Deng18}.
The Hamiltonian for these two degrees of freedom takes the form
\begin{equation}
    H = 
    \EM i \gamma_1 \gamma_2 + \delta_\R d^\dagger d
    + \left(
        t_\R d^\dagger \gamma_1
        + \tilde{t}_\R d^\dagger \gamma_2 + \mathrm{h.c.}
    \right),
\end{equation}
where $\gamma_{1,2}$ are Majorana operators for the wire state and $d^{(\dagger)}$ is a fermionic annihilation (creation) operator for the QD level~\footnote{While, for simplicity, the model described does not include the injector QD, we have verified that the results are unaffected when including explicitly the injector QD as long as it is sufficiently detuned ($\delta_\L \gg t_\L, \tilde{t}_\L$), such that its coupling to the wire can be diagonalized using a perturbative Schrieffer--Wolff transformation.}.
This model is sufficiently general to describe a single low energy state which may be present as part of a trivial or topological phase.
In the topological regime, $\gamma_1$ and $\gamma_2$ are MZMs localized at the ends of the wire and the energy $\EM$ relates to their overlap, while ($\tilde{t}_\R$) $t_\R$ describes the (non-)local coupling of the QD to the (far-side) near-side MZM.
In the trivial case, $\gamma_{1,2}$ combine to form a low-energy spin-resolved Andreev state~\cite{Aghaee25b}.
Using linear response theory, we obtain the parity-dependent quantum capacitance response of the readout QD in the limit of small drive frequency and amplitude~\cite{Aghaee25a}
\begin{equation}
    \Cq(P) = \frac{e^2 \alpha^2 | \tP|^2 }{\EP^3} 
    \tanh\left(\frac{\EP}{2\kB T}\right),
    \label{eq:cq_static}
\end{equation}
where $\tP  = t_\R + i P\,\tilde{t}_\R$ is the effective coupling between the QD level and the wire state.
The corresponding eigenenergies are $\pm \EP$ for total parity sector $P = \pm 1$ of the readout QD and wire system with $\EP = [(\delta_\R + 2P\EM)^2 + 4|\tP|^2]^{1/2}$.
As a function of the QD detuning, \Cref{eq:cq_static} is maximal at $\delta_\R = \delta(P)$ where $\delta(P) \equiv -2P\EM$.
By sweeping $\delta_\R$ and controlling $P$ using the injector QD, one can extract the energy of the wire state $\EM = |\delta(-1) - \delta(+1)|/4$.
In addition, one can, in principle, infer additional information about the nature of the wire state from the shape of the quantum capacitance peak in the two parity sectors.
In particular, deep inside the topological phase $\tilde{t}_\R \to 0$, the width and amplitude of the response becomes parity-independent. In contrast, for a generic state, the peak width and amplitude can exhibit a pronounced parity dependence.
Here we focus on a single state in the wire, but the case of multiple states (quasi-MZMs) can be considered in the same manner as in Ref.~\onlinecite{Aghaee25a} while the gapless phase can be considered numerically as in Ref.~\onlinecite{Boutin25}.
In both of these non-topological cases, we expect, based on these previous results, that the parity dependence of the signal will be strongly suppressed away from very fine-tuned points as the local parity sensor becomes independent of the parity injection on the other side of the wire.

Using the single-state model introduced above, we define a spectroscopy measurement to probe the wire state energy.
The core measurement consists of finely sweeping the plunger gate of the readout QD to probe a charge transition where the QD becomes resonant with the wire state.
At each point of the sweep, the rf response is measured while the plunger voltage of the injector QD is stepped so that it either injects or removes a single electron from the wire.
Here we define the signals $\SA$ and $\SB$, measured in the two parity configurations $A$ and $B$ of the injector, as the complex rf voltage that is reflected from the readout resonator projected onto the dominant quadrature that captures the change in response controlled by the readout QD detuning.
Similarly, the noise $\sigma$ is defined as the standard deviation in the orthogonal quadrature.
This core measurement can then be repeated over a range of wire electrochemical potentials (which we control via the wire-plunger gate $\Vwp$).
The outcome of this measurement is shown in \Cref{fig:em_spectroscopy}(c,d) for the case where QD2 is the readout QD and QD1 the injector QD.
In this experiment, the wire-plunger voltage is swept over a range of \SI{7}{\mV} in the vicinity of the proximitized NW depletion point.
Panel (c) shows the rf signal averaged over the injector states, while in panel (d) the difference in signal is shown.

\begin{figure}
\includegraphics[width=8.64cm]{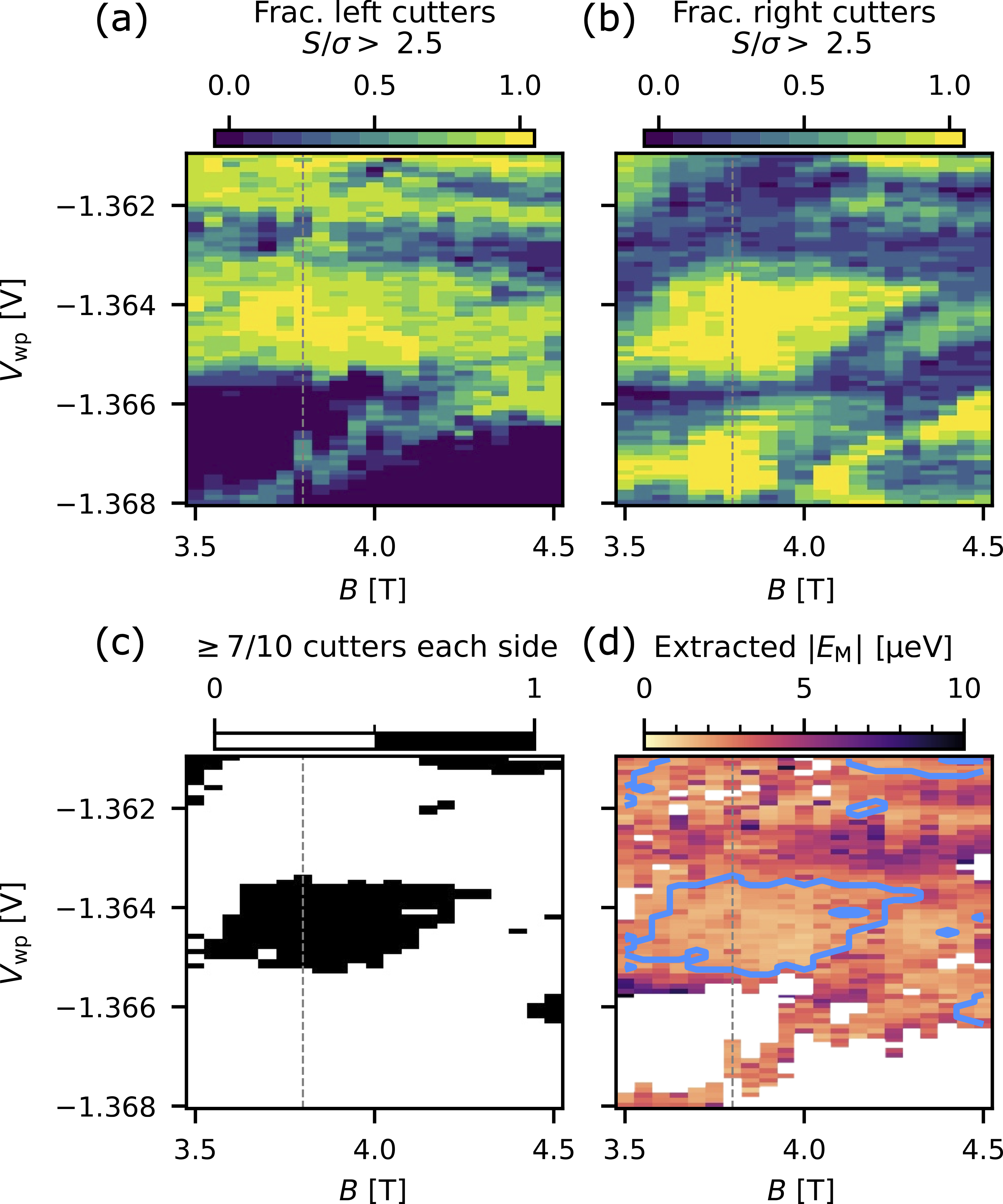}
\caption{
\textbf{Regions with a stable low-energy state at both ends of the wire detected by rf-based wire tuning.}
(a,b)~Fraction of measured left (right) wire cutter configurations where the signal measured on the left (right) QD is above the selected threshold $S / \sigma > 2.5$ where $S$ is the measured rf signal along the quadrature with maximal signal and $\sigma$ the noise along the orthogonal quadrature.
(c)~Regions where the measured signal at both ends of the wire is above threshold for at least 7/10 cutter configurations.
(d)~Estimated energy $\EM$ of the wire low-energy mode averaged over the value extracted from the measurements at each end of the wire.
Overlaid in blue is the contour of the region of interest of panel (c).
In all panels, the dashed grey vertical line indicate the magnetic field (parallel to the wire) $B = \SI{3.8}{\tesla}$ where the data of \Cref{fig:em_spectroscopy} was taken.
}
\label{fig:roi}
\end{figure}

Repeating the measurement with the role of QD1 and QD2 reversed leads to \Cref{fig:em_spectroscopy}(e,f).
In each panel, multiple charge transitions of the readout QD are visible over a range of wire-plunger voltages where they couple to a low-energy wire state.
In regions where the wire state energy $\EM$ is sizable, the rf response splits into a double peak structure, with each branch corresponding to a different parity.
This can be confirmed in \Cref{fig:em_spectroscopy}(d,f), where the branches have opposite signs meaning that they are measured for different occupation of the injector QD.
Additional sign switches are interpreted to be the result of external quasiparticle poisoning events (cf. \Cref{sec:zmpr}).
Within the single-state model presented here, the injector-averaged response [\Cref{fig:em_spectroscopy}(c,e)] is a measure of how well the wire state couples to the readout QD.
We interpret regions of low visibility that persist for different readout QD states as an indication that the wire state has very little weight close to the end that is probed by the readout QD.
When the response in \Cref{fig:em_spectroscopy}(c,e) is sizable, the parity difference [\Cref{fig:em_spectroscopy}(d,f)] indicate states that span the full wire and have measurable $\EM$.
In the topological regime with large $\DeltaT$, one expects a strong injector-averaged signal, while the parity contrast remains weak.

\Longfigref{fig:em_spectroscopy}(g) presents the wire state energy $\EM$ extracted from the parity-dependent shift of the rf response.
To increase statistics, $\EM$ is extracted by estimating the parity-dependent shift of up to 15 QD charge transitions.
From the lever arm of the QD plunger gate and the finite resolution of the DAC controlling the gate voltage, we expect the resolution of this analysis to be limited to $\sim \SI{1}{\ueV}$.
The shaded gray area indicates the region of low-energy wire states identified below.
As the measurements in \Cref{fig:em_spectroscopy} and \Cref{fig:roi} were taken several days apart some small shifts in gate voltages are expected~\footnote{Based on the position of features, we estimate the region with a stable low-energy state shifted to more positive voltage by approximately $\SI{200}{\uV}$ in the interval between the measurements.}.
Notably, the resolution of $\sim \SI{1}{\ueV}$ significantly exceeds that of conductance measurements which can resolve $\EM$ of about the half-width-half-max of a temperature broadened conductance peak $\EM \sim 1.76 \kB T \approx \SI{7.6}{\ueV}$  for $T = \SI{50}{\mK}$.
The high resolution, combined with the ability to resolve left-right correlations of wire-end states even at finite energies, enables the observation of small-amplitude Majorana oscillations~\cite{Cheng09, DasSarma12}.

We now use this type of spectroscopy measurement to identify regions of parameter space where low-energy wire states are stable to local variations.
To this end, we expand the measurement by scanning both the magnetic field and the voltage applied to the wire cutter adjacent to the readout QD over a range of \SI{80}{\mV}.
This provides a 6-dimensional dataset that can be analyzed to identify regions of stable low-energy states in the $(\Vwp, B_z)$ parameter space.

For each configuration, the QD detuning axes are processed to reduce the dataset dimensionality.
We first take the minimum over the injector QD detuning and, second, we extract the maximum
of the signal along the readout QD axis.
By taking the minimum along the injector QD detuning axis first, the procedure selects for low-energy modes which are expected to show sizable signal (at a fixed parameter point) independent of parity changes.
Overall, this procedure yields an effective representation of the local response of low energy states that is insensitive to the precise choice of QD detuning and robust against fluctuations in tunnel coupling between QD and NW for different QD charge transitions.

To identify regions associated with stable low-energy states, we evaluate, for each point in $\Vwp$ and $B_z$, the fraction of wire-cutter configurations for which the rf response exceeds a signal-to-noise ratio $S / \sigma > 2.5$.
This is shown in \Cref{fig:roi}, where \Cref{fig:roi}(a) presents the aggregation of measurements on QD1 and \Cref{fig:roi}(b) on QD2.
\Longfigref{fig:roi}(c) shows the resulting aggregated maps where both datasets show above threshold signal in at least $7/10$ of the measured wire cutter configurations.
The overlap between the left and right junction maps reveals well-defined clusters in magnetic-field and wire-plunger space, corresponding to regions where both ends of the NW exhibit a consistent low-energy response.
These correlated regions are evidence of stable zero-energy states that are simultaneously accessible from both ends of the wire.
While the exact boundary of the identified region is in general sensitive to the details of the thresholds used in the above analysis, the qualitative features are robust to variations of those parameters.

\Longfigref{fig:roi}(d) shows the evolution of $\EM$ across parameter space as a function of the wire-plunger gate voltage and the in-plane magnetic field.
We observe that throughout the identified region of interest $\EM$ is on the order of the measurement resolution $\approx \SI{1}{\ueV}$, which confirms the presence of an extended region with stable low-energy modes.
At the boundaries of this region, the zero-bias peaks split, giving rise to the darker features that surround the identified cluster.

The rf measurements presented here provide a natural and effective tuning step for tetron devices prior to parity readout measurements.
They enable the identification of correlated low-energy states in the NW that exhibit minimal energy splitting, thereby defining favorable operating points for qubit operation.
We emphasize that our rf tuning protocol is distinct from and complementary to the TGP \cite{Pikulin21, Aghaee23}.
While the TGP enables mapping the large-scale phase diagram of simpler devices via DC transport measurements, our rf tuning protocol is used to identify regions of small $\EM$ for qubit operations.

\begin{figure*}
\includegraphics[width=\textwidth]{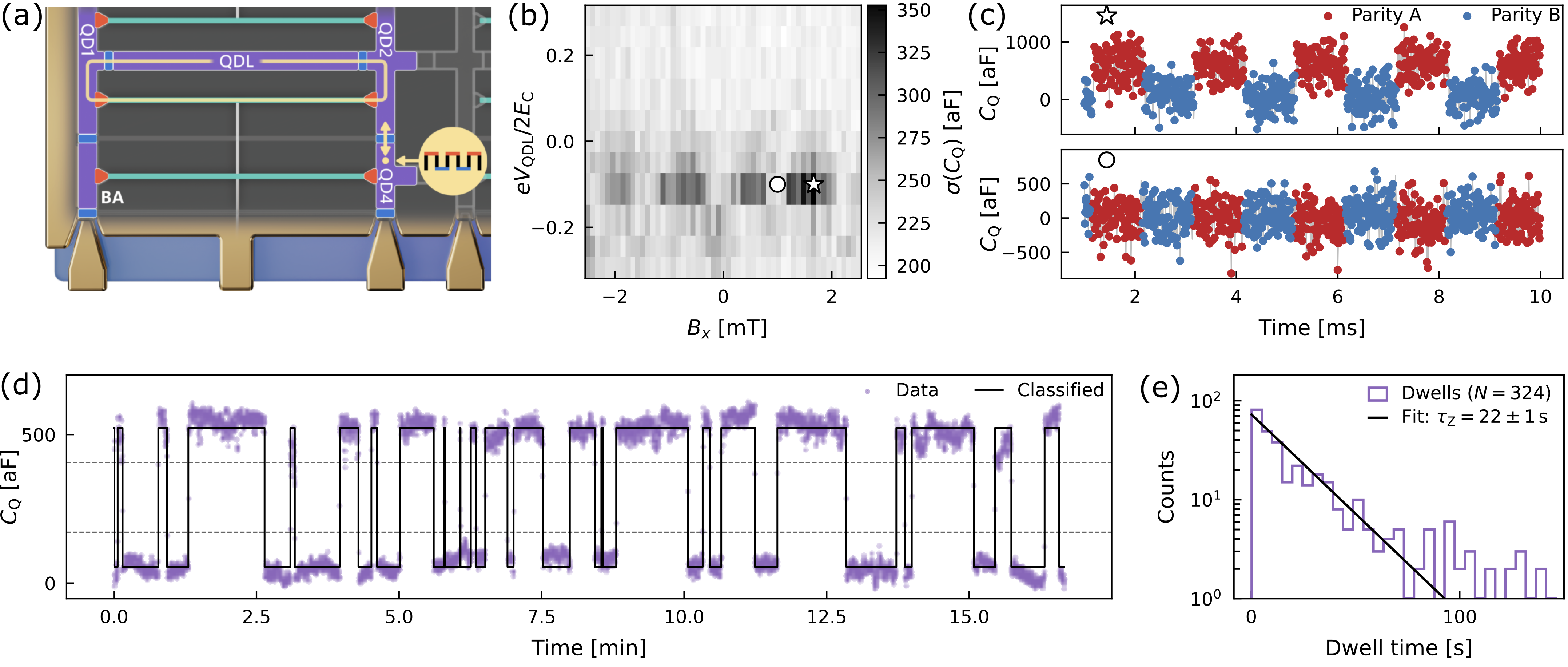}
\caption{
\textbf{$h/2e$-periodic bimodality. Long time traces.}
(a)~Schematic of the Z measurement loop with parity injection.
Depending on the detuning of a nearby QD4 an electron is injected (red) into the Z measurement loop or remains on the QD (blue).
(b)~Standard deviation of time traces, in the presence of parity injection pulses via QD4, measured on QDL as a function of $B_x$ and detuning of QDL.
The parity jumps due to parity injection pulses leading to a flux-periodic increase in the standard deviation.
Flux values of high (low) standard deviation are marked by a star (circle).
(c)~Full time traces of the $\Cq$ signal measured over $\SI{10}{\us}$ integration time at the parameter points marked by a star (circle) in panel (b).
The step function pulses on QD4 with period $\SI{2}{\ms}$ (indicated by blue/red coloring) manifests as periodic jumps in the $\Cq$ response with flux-dependent visibility.
(d)~Example of a long time trace at $\SI{200}{\ms}$ integration time with parity injection pulses turned off at the same parameter point as (c)$\star$.
The data is classified into two states based on thresholding with the state switching whenever the signal exceeds (drops below) the high (low) threshold indicated by the dashed lines.
(e)~Dwell time analysis of the classified time trace in (d) and 8 additional time traces (taken from repeat measurements and ranges of $\Vwp = \SI{-1.36513}{\volt} \pm \SI{10}{\uV}$) showing a fit to an exponential dwell time distribution with characteristic time scale of $\tauZ = 22 \pm \SI{1}{\second}$.
}
\label{fig:time_traces}
\end{figure*}

\section{Interferometric parity readout}
\label{sec:zmpr}

The Pauli-$Z$ measurement is a fundamental operation of a tetron.
It can be implemented by probing the parity of the NW via an interference loop formed between the NW and a long QD.
This QD is coupled to the ends of the NWs through smaller intermediary QDs.
As demonstrated in previous works \cite{Aghaee25a, Aghaee25b}, the signature of this measurement is an oscillatory bimodality of the quantum capacitance ($\Cq$) response as a function of magnetic flux, with a periodicity determined by the enclosed area of the interference loop.

At flux values corresponding to maximal interference visibility, the $\Cq$ signal exhibits a telegraph-like time dependence, characterized by stochastic switching events governed by quasiparticle poisoning dynamics.
When quasiparticle poisoning times are sufficiently long, an alternative approach can be employed: controlled parity switching through electron injection.
Specifically, an electron can be injected into the NW from a QD that is not part of the interferometric loop, thereby toggling the parity of the system.
This process is equivalent to a quasiparticle relaxing into the lowest-energy mode of the NW.

\Longfigref{fig:time_traces}(a) illustrates a schematic of the device, highlighting the injection mechanism.
In this configuration, quantum dot 4 (QD4) is decoupled from the bottom NW by closing the corresponding wire cutter, while remaining weakly coupled to the interferometric loop via gate DC24.
Measurements are performed following similar procedures as in previous works \cite{Aghaee25a, Aghaee25b}, recording time traces of $\Cq$ as a function of flux and QD detuning within the interferometric loop.
In addition, a square voltage pulse is applied to QD4 during the measurement, with an amplitude corresponding to one electron charge (pulsing from one Coulomb valley to the next) and a frequency of $\SI{0.5}{\kHz}$.
The wire-plunger voltage $\Vwp = \SI{-1.36513}{\volt}$ is tuned to a regime where $\EM$ is minimized which is consistent with the low energy region determined in \Cref{sec:spectroscopy} up to minor shifts.

\Longfigref{fig:time_traces}(b) presents the standard deviation $\sigma(\Cq)$ of the $\Cq$ time traces as a function of magnetic flux and QDL gate voltage.
Here, the flux is tuned by varying the $B_x$ component of the external magnetic field.
Due to a small misalignment between the device and the external magnet, $B_x$ does not correspond exactly to the out-of-plane field component ($B_{\perp}$) that determines the flux, resulting in an offset of approximately $\SI{-10}{\mT}$ that is not corrected for in the data shown here.
A pronounced flux dependence is observed when QDL is tuned close to resonance, with a periodicity of $1.3 \pm \SI{0.3}{\mT}$.
This is consistent with $h/2e$ periodicity, which corresponds to a magnetic field period of $1.0 \pm \SI{0.3}{\mT}$, determined by the area of the loop.
\Longfigref{fig:time_traces}(c) shows representative time traces of $\Cq$ corresponding to flux points of maximal (minimal) standard deviation $\sigma(\Cq)$.
At flux values of high $\sigma(\Cq)$, the time traces exhibit a clear step function behavior, with switching events occurring at the applied injection rate of $\SI{0.5}{\kHz}$, while they are absent for the in-between flux values of low $\sigma(\Cq)$, as expected.

This measurement technique enables rapid identification of optimal operating points for interferometric measurements, without being limited by the intrinsic parity switching time associated with quasiparticle poisoning.
At the same parity-sensitive operating point, we perform additional measurements with the voltage pulse on QD4 disabled.
In this configuration, the time trace of $\Cq$ reveals the intrinsic $Z$-parity lifetime of the device which far exceeds the measurement time of the pulsed time traces, see \Cref{fig:time_traces}(d).
To quantify the lifetime we classify the data by fitting a Gaussian mixture model which estimates the center values of two states $\Cqh$ and $\Cql$ and sets a high (low) threshold of $0.75 \Cqh + 0.25 \Cql$.
The use of two thresholds avoids accidental misclassification due to possible drift during the long time traces.
By aggregating multiple measurements, we observe a total of $N = 324$ dwell intervals.
The data are consistent with a single exponential distribution, as expected for a homogeneous Poisson process due to high-energy quasiparticle poisoning.
We extract a characteristic parity lifetime and corresponding fit uncertainty of $\tauZ = 22 \pm \SI{1}{\second}$ by performing an exponential fit on the measured dwell time distribution as shown in \Cref{fig:time_traces}(e).

We attribute the more than 3 orders of magnitude enhancement in parity lifetime, relative to Refs.~\onlinecite{Aghaee25a, Aghaee25b}, to the difference between Pb and Al.
It is more difficult to break Cooper pairs since our epitaxial Pb films have higher gap $\approx \SI{1300}{\ueV}$ than Al $\approx \SI{300}{\ueV}$, thereby suppressing quasiparticle generation \cite{Wilson04, Aumentado04, Barends11}.
Quasiparticles are also more likely to recombine quickly due to the stronger electron-phonon coupling.
These effects reduce poisoning events and extend the stability of the parity state.
The change to GaSb substrate and introduction of Sb in the quantum well were both designed to increase the topological gap, but they have subleading effect on the non-equilibrium quasiparticle density.

\section{Discussion and outlook}
\label{sec:discussion}

Our results confirm a central premise of topological quantum computing: increasing the excitation gap dramatically reduces error mechanisms and improves qubit performance.
By incorporating a higher-$\Tc$ superconductor (Pb) and an optimized semiconductor heterostructure into a multi-tetron array, we achieve a robust topological phase with a significantly enlarged topological gap and an extended topological phase region [see \Cref{fig:material_stack2}].
These material and design improvements directly yield qubit-level benefits: we observe Majorana parity lifetimes exceeding $\SI{20}{\second}$, over three orders of magnitude longer than those in comparable Al-based tetrons (Refs.~\onlinecite{Aghaee25a, Aghaee25b}).
Qubit operations are no longer limited by non-equilibrium quasiparticles on experimental timescales.
A parity switching time of $\sim \SI{20}{\second}$ is more than seven orders of magnitude longer than typical qubit operation times ($\sim \SI{1}{\us}$), allowing an extremely large number of operations within a fixed parity state.
We expect that parity lifetimes in floating (charging-energy protected) tetrons could be even longer, since their smaller superconducting islands may further suppress residual quasiparticle effects.
Thus, the probability of any unintended parity flip during a typical qubit operation becomes effectively negligible.

We have also developed a scalable tuning protocol based on dispersive quantum-dot readout, enabling parallel characterization of multiple NWs.
Using this approach, we directly probe low-energy states at both ends of a NW and extract the Majorana splitting energy $\EM$.
Our parity-switching spectroscopy confirms that $\EM$ is below our $\sim \SI{1}{\ueV}$ resolution across extended parameter regimes, in line with design simulations and smaller than typical values in earlier Al-based devices.
Within the relevant operating range, therefore, both key error channels---Majorana hybridization and quasiparticle poisoning---are strongly suppressed.

Looking ahead, the values of $\EM$ achieved here are expected to have beneficial implications for Pauli-$X$ measurements, whose characteristic switching time $\tauX$ scales as $\EM^2$ \cite{Aghaee25b}.
Deep in the topological regime, $\EM \sim \DeltaT \exp(-L/\xiT)$, where $\xiT$ is the topological coherence length. Increased
$\DeltaT$ of the type demonstrated here leads to shorter $\xiT$
(in the perfectly clean case, $\xiT = \vF / \DeltaT$, where $\vF$ is the Fermi velocity, but the relation is more complicated in the generic case).
Increasing the NW length $L$ exponentially suppresses $\EM$ and thus dramatically extends $\tauX$,
which suggests that $\tauX$ could be more than an order of magnitude longer than in previous devices.
Investigating these effects will be an important direction for future work.

Finally, our demonstration constitutes a significant step towards scalable, fault-tolerant Majorana-based quantum computing.
The multi-tetron array presented here functions as a modular ``unit cell'' for a larger architecture; it can be tiled into much larger qubit arrays (e.g., a 12-qubit array) without altering the underlying control or readout approach.
The strong parity protection observed in our tetron prototype suggests that, even as the system scales up, each qubit will remain well isolated from non-equilibrium quasiparticles and will benefit from exponentially suppressed error rates due to the large $\DeltaT$.
The primary challenge for scaling will be to preserve and further improve these favorable conditions in larger devices and across broader parameter ranges.
Continued refinements in materials quality and device design---including driving $\EM$ to zero and eliminating any residual nonequilibrium quasiparticles---will be crucial for realizing high-fidelity, fault-tolerant multi-qubit Majorana arrays.

\begin{acknowledgments}
We thank Gabriel Aeppli, Peter Armitage, Haim Beidenkopf, and Jacob R. Taylor for discussions and Pureum Lee and Edward Lee for assistance with the figures.
\end{acknowledgments}

\vspace{1cm}
\textbf{Correspondence and requests for materials} should be addressed to Chetan Nayak (cnayak@microsoft.com).

\vspace{1cm}
$^\dagger${\small
Morteza Aghaee, Zulfi Alam, Mariusz Andrzejczuk, Andrey Antipov, Theodora Asimakidis, Mikhail Astafev, Lukas Avilovas, Ahmad Azizimanesh, Amin Barzegar, Bela Bauer, Jonathan Becker, Umesh Kumar Bhaskar, Andrea G. Boa, Srini Boddapati, Nichlaus Bohac, Jouri Bommer, Jan Borovsky, L\'{e}o Bourdet, Samuel Boutin, Srivatsa Chakravarthi, Benjamin J. Chapman, Nikolaos Chatzaras, Tzu-Chiao Chien, Jason Cho, Patrick T. Codd, William Cole, Paul W. Cooper, Fabiano Corsetti, Ajuan Cui, Tareq El Dandachi, Konstantinos Divanis, Clayton Doyle, Andreas Ekefjard, Javier A. Falcon, Saeed Fallahi, Luca Galletti, Geoffrey C. Gardner, Haris Gavranovic, Jo\~{a}o Pedro Morais Gomes, Deshan Govender, Flavio Griggio, Ruben Grigoryan, Sebastian Grijalva, Sergei Gronin, Jan Gukelberger, Marzie Hamdast, Esben Bork Hansen, Sebastian Heedt, Samantha Ho, Laurens Holgaard, Kevin van Hoogdalem, Jinnapat Indrapiromkul, Henrik Ingerslev, Lovro Ivancevic, Max Jantos, Thomas Jensen, Jaspreet Singh Jhoja, Vidul R. Joshi, Konstantin V. Kalashnikov, Ray Kallaher, Rachpon Kalra, Farhad Karimi, Torsten Karzig, Maren Elisabeth Kloster, Christina Knapp, Jonathan Knoblauch, Jonne Koski, Anders Kringh\o{}j, Tom Laeven, Jeffrey Lai, Gijs de Lange, Thorvald W. Larsen, Kyunghoon Lee, Kongyi Li, Shuang Liang, Tyler Lindemann, Luna Lochmatter, Marijn Lucas, Roman Lutchyn, Morten Hannibal Madsen, Nasiari Madulid, Ivan Maliyov, Yanick Mampaey, Michael Manfra, Signe Brynold Markussen, Esteban A. Martinez, J. R. Mattinson, M\'{o}nica Meira, Camille A. Mikolas, Sarang Mittal, Gopakumar Mohandas, Christian Mollgaard, Michiel W. A. de Moor, Chris Moore, George Moussa, Bhargav Nabar, Anirudh Narla, Ahmad Naseri, Chetan Nayak, Bj\o{}rn Funch Schr\o{}der Nielsen, Jens Hedegaard Nielsen, Michael J. Nystrom, Eoin O'Farrell, Keita Ohtani, Theodore Rex Orth, Sara Di Paolo, Camille Papon, Luca Petit, Dima Pikulin, Athanasia Piperidou, Mohana Rajpalke, Alejandro Alcaraz Ramirez, Katrine Rasmussen, David Razmadze, Yuan Ren, Mariya Romanova, Lo\"{\i}c Roure, Ivan Sadovskyy, Lauri Sainiemi, Juan Carlos Estrada Salda\~{n}a, Irene Sanlorenzo, Tatiane Pereira dos Santos, Carl Vincent C. Saruda, Simon Schaal, John Schack, Emma R. Schmidgall, Christina Sfetsou, Cristina Sfiligoj, Zahra Shekason, Sarat Shankar Sinha, Patrick Sohr, Maria Jo\~{a}o Louren\c{c}o de Sousa, Kasper Roed Spiegelhauer, Tomas Stankevic, Henri J. Suominen, Judith Suter, Attila Sz\'{e}n\'{a}si, Samuel M. L. Teicher, Naganivetha Thiyagarajah, Raj Tholapi, Mason Thomas, Dennis Tom, Emily Toomey, Joshua Tracy, Michelle Turley, Matthew D. Turner, Ivan Urban, Aakash Valliappan, Dmitrii V. Viazmitinov, Anna Wulff Viazmitinova, Dominik Johannes Vogel, Wenbo Wang, Christopher A. Watson, John Watson, Alex Webster, Joseph Weston, Timothy Williamson, Georg W. Winkler, David J. van Woerkom, Brian Paquelet Wuetz, C\'{e}cile X. Yu, Emrah Yucelen, Jes\'{u}s Herranz Zamorano, Roland Zeisel, Guoji Zheng, A. M. Zimmerman.
}

\bibliography{pb_zmpr}

\end{document}